\documentclass{article} 
\usepackage{spconf,amsmath,amssymb,graphicx}
\usepackage{mathtools}
\usepackage[caption=false,font=footnotesize]{subfig}
\usepackage{algpseudocode}
\usepackage{algorithm}
\usepackage{color}
\usepackage{bm}
\usepackage{amsfonts}
\usepackage{amscd}
\usepackage[mathcal]{eucal}
\usepackage{epsfig}
\usepackage{caption}
\usepackage{array}
\usepackage{eqparbox}
\usepackage{enumerate}
\usepackage{stackrel}
\usepackage{pdflscape}
\usepackage{lineno}

\newcommand{\ie}{{\em i.e.}}

\newcommand{\iid}{i.i.d.}
\newcommand{\apriori}{{\em a priori}}

\newcommand{\secref}[1]{Section~\ref{#1}}
\newcommand{\figref}[1]{Fig.~\ref{#1}}

\newcommand{\qed}{\nobreak \ifvmode \relax \else
      \ifdim\lastskip<1.5em \hskip-\lastskip
      \hskip1.5em plus0em minus0.5em \fi \nobreak
      \vrule height0.75em width0.5em depth0.25em\fi}

\title{SEQUENTIAL JOINT SIGNAL DETECTION AND SIGNAL-TO-NOISE RATIO ESTIMATION}
\name{M. Fau{\ss}$^{\ast}$, K. G. Nagananda$^{\dag}$, A. M. Zoubir$^{\ast}$, and H. V. Poor$^{\ddag}$\thanks{This research was supported in part by the U. S. National Science Foundation under Grants CNS-1456793 and ECCS-1343210.}}
\address{$^\ast$ Signal Processing Group, Darmstadt University of Technology, D-64283 Darmstadt, Germany \\ $^\dag$ Dept. of Electronics and Communications Engineering, PES University, Bangalore 560085, India \\ $^\ddag$ Dept. of Electrical Engineering, Princeton University, Princeton, NJ 08544, USA}

\begin{document}
\ninept
\maketitle

\begin{abstract}
The sequential analysis of the problem of joint signal detection and signal-to-noise ratio (SNR) estimation for a
linear Gaussian observation model is considered. The problem is posed as an optimization setup where the goal is to
minimize the number of samples required to achieve the desired (i) type I and type II error probabilities and (ii) mean
squared error performance. This optimization problem is reduced to a more tractable formulation by transforming the
observed signal and noise sequences to a single sequence of Bernoulli random variables; joint detection and estimation
is then performed on the Bernoulli sequence. This transformation renders the problem easily solvable, and results in a
computationally simpler sufficient statistic compared to the one based on the (untransformed) observation sequences.
Experimental results demonstrate the advantages of the proposed method, making it feasible for applications having
strict constraints on data storage and computation.
\end{abstract}

\vspace{-0.1in}
\section{Introduction} \label{sec:introduction}\vspace{-0.1in}
The joint problem of distinguishing between different hypotheses and estimating the unknown parameters based on the
outcome of the hypotheses test has received considerable attention in the literature \cite{Middleton1968} -
\nocite{Fredriksen1972}\nocite{Moustakides2012}\nocite{Chen2013}\nocite{Li2015}\cite{Li2016}. Such a problem arises in
a wide range of applications, including (i) radiographic inspection for detecting anomalies in manufactured
objects and estimating their position and size \cite{Fillatre2007}, (ii) retrospective changepoint hypotheses testing
to detect change in the statistics and simultaneously estimate the time of change \cite{Vexler2009},
\cite{Boutoille2010}, (iii) jointly detecting the presence of multiple objects and estimating their states using image
observations \cite{Vo2010}, and (iv) distinguishing between two hypotheses and at the same time estimating the unknown
parameters in the accepted hypothesis in a distributed framework \cite{Zhu2016}. Some popular techniques to address
this problem include reformulating the composite detection problem as a pure estimation problem \cite{Ghobadzadeh2012},
while the maximum a posteriori estimate was shown to provide a solution to the joint detection and estimation problem
in a Bayesian context \cite{Gelman2013}. The problem has also been addressed in a sequential setting, where the
objective is to minimize the number of samples subject to a constraint on the combined detection and estimation cost
\cite{Yilmaz2015}, \cite{Yilmaz2016}. The generalized sequential probability ratio test was presented in \cite{Li2014},
where a decision was obtained using the maximum likelihood estimate of the unknown parameter.

There is another class of problems where it is desirable to distinguish between the hypotheses and simultaneously
estimate the signal-to-noise ratio (SNR), specifically signal and noise powers, under the ``signal present''
hypothesis. For example, in speech processing, it was shown in \cite{Sohn1998}, \cite{Sohn1999} that the performance of
voice detection systems can be drastically improved by jointly estimating the noise power and {\apriori} SNR. In
\cite{Le2013}, it was shown that a scheduling scheme performed detection in an energy-efficient manner by jointly
estimating the SNR. However, \cite{Le2014} reported that the techniques developed in some of the papers mentioned above
were not readily applicable to the problem of joint detection and signal and noise power estimation. For a Bayesian
formulation, it was shown \cite[Sec. III]{Le2014} that knowledge of the priors $\text{Pr}(H_0)$ and $\text{Pr}(H_1)$ or
of the distribution of the unknown parameters was not amenable for problems addressed in \cite{Sohn1998} -
\nocite{Sohn1999}\cite{Le2013}. Instead, an optimal solution for Gaussian observation models was presented using
conjugate priors on the signal and noise powers \cite[Sec. IV]{Le2014}.

In this paper, we extend the problem of joint signal detection and SNR estimation, without {\apriori} knowledge of the
signal or noise powers, to a \emph{sequential setting} and propose a novel method to address this problem. To the best
of our knowledge, the sequential analysis of this problem has not been reported in the literature.  The problem of
distinguishing between two hypotheses (signal absent and signal present) and at the same time estimating the SNR in a
Gaussian observation model is posed as an optimization setup, where we seek to minimize the number of samples required
to achieve the desired (i) type I and type II error probabilities and (ii) mean squared error (MSE) performance. Our
approach comprises transforming the observed signal and noise sequences to a single sequence of Bernoulli random
variables, and then performing the detection-and-estimation task on the resulting Bernoulli sequence.

One of the main advantages of this transformation is that it significantly reduces the complexity of the optimization
problem so that it can be solved more efficiently. Secondly, we obtain a computationally simpler sufficient statistic
compared to the one that emerges when solving the problem directly. Moreover, we show that the proposed method allows
for more degrees of freedom than an equivalent Bayesian solution.
Experimental results show that (i) the expected number of measurements required to achieve the desired
performance almost remains constant for increasing values of SNR, and (ii) many of the constraints in the transformed
optimization problem are inactive which renders the problem easily solvable. As such, the method developed in this
paper is feasible especially for applications with strict constraints on data storage and computation.

In \secref{sec:problem_formulation}, we present the problem statement. In \secref{sec:solution}, we detail the transformation of the observations to a Bernoulli sequence, and show how the original optimization problem can be reformulated into a setup which can be solved efficiently. Results of computer simulations are presented in \secref{sec:experiments}. Concluding remarks are provided in \secref{sec:conclusion}.

\vspace{-0.1in}
\section{Problem Formulation}\label{sec:problem_formulation}\vspace{-0.1in}
The following linear Gaussian signal model is considered:
\begin{eqnarray}
x[n] = s[n] + w[n], \quad n=1, 2, \ldots,
\label{eq:signal_model}
\end{eqnarray}
where $\bm{x} \triangleq \left\{x[1], x[2], \dots \right\}$ denotes the set of observations, while $\bm{s} \triangleq \left\{s[1], s[2], \dots \right\}$ and $\bm{w} \triangleq \left\{w[1], w[2], \dots \right\}$ denote sets of {\iid} zero mean Gaussian random variables corresponding to the signal and noise, respectively. The variances $\sigma_s^2 \geq 0$ and $\sigma_w^2 > 0$ are unknown, and the SNR is given by $\theta = \frac{\sigma_s^2}{\sigma_w^2}$. The problem is to distinguish between the two hypotheses
\begin{eqnarray}
\begin{cases}
H_0: \theta = 0 & \text{(signal absent)}, \\
H_1: \theta \geq \theta_{\min} & \text{(signal present)},
\end{cases}
\label{eq:hypotheses_theta}
\end{eqnarray}
and at the same time estimate the SNR under the hypothesis $H_1$ using as few samples as possible, while satisfying
predefined constraints on the type I and type II error probabilities. In \eqref{eq:hypotheses_theta}, $\theta_{\min}$
denotes the minimum SNR for which reliable detection is to be guaranteed. We attempt to \emph{jointly} solve the
problem of signal detection and SNR estimation in a \emph{sequential} setting. While the latter enables one to adapt
the number of samples to the quality of realizations, the former ensures that the dual objective of detection and
estimation is achieved with a desired performance. Essentially, the problem can be formulated as the following
optimization setup:
\begin{eqnarray}
&& \min_{\psi,\delta,\hat{\theta}} ~ \mathbb{E}_{\theta^*}[N] \label{eq:sequential_theta1} \\
\nonumber \text{subject to} \\
&& P_0(\delta_N = 1) \leq \alpha, \label{eq:sequential_theta2} \\
&& P_{\theta}(\delta_N = 0) \leq \beta(\theta), \quad \forall \theta \geq \theta_{\min}, \label{eq:sequential_theta3} \\
&& \mathbb{E}_{\theta}\left[(\hat{\theta}_N-\theta)^2\right] \leq \gamma(\theta), \quad \forall \theta \geq \theta_{\min},
\label{eq:sequential_theta4}
\end{eqnarray}
where $N$ denotes the sample number at which the sequential test is terminated, $\psi_n, \delta_n \in \{0,1\}$ denote
the stopping and decision rule after the $n^{\text{th}}$ sample has been observed, and $\hat{\theta}_n$ is an estimator
for $\theta$. The constant $\theta^{\ast}$ is a nominal SNR value under which the average sample number (ASN) is to be
minimum. $P_{0}(\cdot)$ and $P_{\theta}(\cdot)$ denote the probabilities of an event under hypothesis $H_0$ and $H_1$,
respectively. The type I and type II error probabilities are bounded by $\alpha$ and $\beta$, respectively, with
$\beta$ being allowed to depend on the true SNR. The mean square error (MSE) of the estimator $\hat{\theta}$ is bounded
by a function $\gamma(\cdot)$.

We assume knowledge of a sequence of noise-only realizations $\tilde{\bm{w}} \triangleq \left[\tilde{w}_1, \tilde{w}_2,
\dots \right]$, that can either be recorded before performing the test, or can be generated on the fly, for example,
via an identical sensor that is shielded from the external signal, but otherwise exposed to the same environmental
conditions. Without $\tilde{\bm{w}}$, the testing problem cannot be solved for the setup considered in this paper.

\vspace{-0.1in}
\section{Solution Methodology}\label{sec:solution}\vspace{-0.1in}
Our approach comprises the following two steps: (i) the two sequences $\bm{x}$ and $\tilde{\bm{w}}$ are transformed to a single sequence of Bernoulli random variables, whose success probability is determined by the \emph{true} SNR, and (ii) a sequential joint detection and estimation procedure is applied to this Bernoulli sequence.
\vspace{-0.1in}
\subsection{Transformation to a Bernoulli sequence}\vspace{-0.1in}
The two Gaussian sequences $\bm{x}$ and $\tilde{\bm{w}}$ are transformed into a single Bernoulli sequence $\bm{b}$
using Birnbaum's sequential procedure \cite{Birnbaum1958} as follows: At every time step, we calculate the sum of the
squares of the samples from both sequences and take an additional sample from the one whose sum is smaller. Whenever
the additional sample changes the order of the two sums, the procedure outputs a 0 or 1 depending on which sequence the
sample was drawn from. Essentially, given $\bm{x}$ and $\tilde{\bm{w}}$, we define $x_s[k] \triangleq
\sum_{n=1}^{k}x[n]^2$ and $\tilde{w}_s[k] \triangleq \sum_{n=1}^{k}\tilde{w}[n]^2$, and $\bm{y} \triangleq \left\{y[1],
y[2], \dots \right\}$, where $y[k] = \min \left\{x_s[k], \tilde{w}_s[k]\right\}$. In the sequence $\bm{y}$, let the
$i^{\text{th}}$ transition from $x_s$ to $\tilde{w}_s$, or vice versa, occur at the $k_i^{\text{th}}$ sample. Then the
Bernoulli sequence is $\bm{b} \triangleq \left\{b[1], b[2], \dots \right\}$, where
\begin{eqnarray}
b[m] =
\begin{cases}
0,~\text{if}~ \tilde{w}_s[k_m] < x_s[k_m],\\
1,~\text{otherwise}.
\end{cases}
\label{def:birnbaum}
\end{eqnarray}
In \cite{Birnbaum1958}, it was proved that irrespective of the actual values of $\sigma_s^2$ and $\sigma_w^2$, the output $\bm{b}$ is a sequence of {\iid} Bernoulli random variables with success probability $\rho = \frac{1}{\theta + 2}$, thereby establishing a one-to-one correspondence between $\theta$ and $\rho$. Therefore, \eqref{eq:hypotheses_theta} can be re-hypothesized in terms of $\rho$, {\ie},
\begin{equation}
\begin{cases}
H_0: \rho = 0.5 & \text{(signal absent)}, \\
H_1: \rho \leq \rho_{\max} & \text{(signal present)},
\end{cases}
\label{eq:hypotheses_rho}
\end{equation}
where $\rho_{\max} = 1/(\theta_{\min}+2)$. The optimization setup \eqref{eq:sequential_theta1} - \eqref{eq:sequential_theta4} can be reformulated as
\begin{eqnarray}
&& \min_{\psi,\delta,\tilde{\theta}}~\mathbb{E}_{\rho^*}[M] \label{eq:sequential_rho1} \\
\nonumber \text{subject to} \\
&& P_0(\delta_M = 1) \leq \alpha, \label{eq:sequential_rho2} \\
&& P_{\rho}(\delta_M = 0) \leq \beta(\rho), \quad \forall \rho \leq \rho_{\max}, \label{eq:sequential_rho3} \\
&& \mathbb{E}_{\rho}\left[(\tilde{\theta}_M-\tfrac{1}{\rho})^2\right] \leq \gamma(\rho), \quad \forall \rho \leq \rho_{\max},
\label{eq:sequential_rho4}
\end{eqnarray}
where $\tilde{\theta} = \hat{\theta}-2$ is an estimator for $\theta$ that is biased in order to simplify the expression
under the expected value. $P_{0}(\cdot)$ and $P_{\rho}(\cdot)$ denote the probabilities of an event under hypothesis
$H_0$ and $H_1$, respectively, corresponding to \eqref{eq:hypotheses_rho}. $M$ denotes the sample number of $\bm{b}$ at
which the procedure is terminated. Since it takes multiple observations of $\bm{x}$ and $\tilde{\bm{w}}$ to generate
one observation of $\bm{b}$, $N$ is in general larger than $M$. Moreover, $N$ includes observations of $\tilde{\bm{w}}$
as well as $\bm{x}$. However, by formulating the problem in terms of $\bm{b}$ and $\rho$, we aim at minimizing the
required samples of $\bm{b}$, irrespective of the number of observations of $\bm{x}$ and $\tilde{\bm{w}}$ are used to
generate these samples. Due to this difference in the objective, both problem formulations are not strictly equivalent
so that the proposed procedure cannot be guaranteed to be strictly optimal. A more detailed analysis of the loss
incurred by applying Birnbaum's transformation is a subject of future research. Also note that, depending on the cost
involved in sampling from $\bm{x}$ and $\tilde{\bm{w}}$, one can modify the transformation to require more or fewer
samples from a certain sequence. This additional potential for optimization is not taken into account in this work.
For more details on the transformation and its near-optimality properties see \cite{Birnbaum1958}.
\vspace{-0.1in}
\subsection{Joint detection and estimation}\label{sec:typestyle}\vspace{-0.1in}
In order to solve \eqref{eq:sequential_rho1}, we first calculate its Lagrangian dual. For fixed Lagrange multipliers, it results in an unconstrained optimal stopping problem that can be solved by means of dynamic programming. We then choose the Lagrange multipliers such that the procedure satisfies the constraints on the error probabilities and the desired estimation accuracy. For analytical tractability, we relax the constraints under $H_1$ to hold for all $\rho \in \mathcal{P}$, where $\mathcal{P} \triangleq \{\rho_1,\ldots, \rho_K\}$ is a discrete subset of $[0,\rho_{\text{max}}]$. That is, we bound $P_{\rho_k}(d = 0)$ and $\mathbb{E}_{\rho_k}[ (\tilde{\theta} - \tfrac{1}{\rho})^2 ]$ only at a finite number of grid points; $\beta(\theta)$ and $\gamma(\theta)$ will, therefore, be approximated for points in-between. For the problem considered in this work, these approximations are shown to be reasonably accurate (see examples in \secref{sec:experiments}).

The Lagrangian dual problem of \eqref{eq:sequential_rho1}, with $[0,\rho_{\text{min}}]$ replaced by $\mathcal{P}$ and Lagrange multipliers $\lambda \in \mathbb{R}^{K+1}$ and $\mu \in \mathbb{R}^K$, is given by
\begin{eqnarray}
\max_{\lambda,\mu \geq 0} \left\{ L(\lambda,\mu) - \lambda_0 \alpha
- \sum_{k=1}^K \left( \lambda_k \beta(\rho_k) + \mu_k \gamma(\rho_k) \right) \right\},
\label{eq:lagrange_dual}
\end{eqnarray}\vspace{-0.1in}
\begin{multline}
L(\lambda,\mu) = \min_{\psi,\delta,\tilde{\theta}} \;
\Biggl\{ \mathbb{E}_{\rho^*}[M] + \lambda_0 P_0(\delta_M = 1) \\
+ \sum_{k=1}^K \left(\lambda_k P_{\rho_k}(\delta_M = 0) +
\mu_k \mathbb{E}_{\rho_k}\left[(\tilde{\theta}_M-\tfrac{1}{\rho_k})^2\right] \right) \Biggr\},
\label{eq:optimal_stopping}
\end{multline}
where $\rho_k \in \mathcal{P}$. Following the techniques developed in \cite{Novikov2009}, \cite{Fauss2015}, \eqref{eq:optimal_stopping} can be straightforwardly solved as follows, where we omit the details in the interest of space: Let $m_0$ and $m_1$ denote the number of 0's and 1's observed. The likelihood-ratios of the corresponding observations under $P_0$ and $P_{\rho_k}$, with respect to $P_{\rho^*}$ are given by
\begin{eqnarray}
Z^{m_0,m_1}_0 &=& \left(\frac{0.5}{1-\rho^*}\right)^{m_0}
              \left(\frac{0.5}{\rho^*}\right)^{m_1}, \\
  Z^{m_0,m_1}_k &=& \left(\frac{1-\rho_k}{1-\rho^*}\right)^{m_0}
                  \left(\frac{\rho_k}{\rho^*}\right)^{m_1}.
\end{eqnarray}
We define, $\forall i \geq 0$, $E^{m_0,m_1}_{\lambda} \triangleq \sum_{k=1}^K \lambda_k Z^{m_0,m_1}_k$ and $E^{m_0,m_1}_{\mu,i} \triangleq \sum_{k=1}^K \rho_k^{-i} \mu_k Z^{m_0,m_1}_k$. The optimal decision rule $\delta^*$ is given by
\begin{eqnarray}
\delta_{m_0, m_1}^* = \begin{cases}
        1, & \lambda_0 Z^{m_0,m_1}_0 \leq E^{m_0,m_1}_{\lambda}, \\[0.5ex]
        0, & \lambda_0 Z^{m_0,m_1}_0 > E^{m_0,m_1}_{\lambda},
     \end{cases}
\label{eq:decision_rule}
\end{eqnarray}
and the optimal estimator $\tilde{\theta}^*$ by
\begin{eqnarray}
\tilde{\theta}_{m_0,m_1}^* = \frac{E^{m_0,m_1}_{\mu,1}}{E^{m_0,m_1}_{\mu,0}}.
\label{eq:optimal_estimator}
\end{eqnarray}
The optimal stopping rule $\psi^*$ is obtained as
\begin{eqnarray}
  \psi_{m_0,m_1}^* =
\begin{cases}
1, & G_{m_0,m_1} = R_{m_0,m_1}, \\
0, & G_{m_0,m_1} > R_{m_0,m_1},
\end{cases}
\label{eq:optimal_stop}
\end{eqnarray}
where
\begin{multline}
  G_{m_0,m_1} = \min \left[\lambda_0 Z^{m_0,m_1}_0 \,,\, E^{m_0,m_1}_{\lambda} \right]
  \\ + E^{m_0,m_1}_{\mu,2} - \frac{(E^{m_0,m_1}_{\mu,1})^2}{E^{m_0,m_1}_{\mu,0}}
  \label{eq:cost_g}
\end{multline}
and $R_{m_0,m_1}$ is defined recursively as follows
\begin{multline}
  R_{m_0,m_1} = \min \left[ G_{m_0,m_1} \right., \\
                \left. 1 + \rho^* R_{m_0,m_1+1}  + (1-\rho^*) R_{m_0+1,m_1} \right].
\end{multline}

The quantities $G_{m_0,m_1}$ and $R_{m_0,m_1}$ correspond to the cost incurred for stopping immediately, or stopping at the optimal time instant, given that $m_0$ 0's and $m_1$ 1's have been observed. The procedure is stopped for the first time when $G_{m_0,m_1} = R_{m_0,m_1}$. The $\min$ term in \eqref{eq:cost_g} signifies the detection cost if the decision rule \eqref{eq:decision_rule} is employed, while the last two terms correspond to the estimation cost, {\ie}, the deviation of the estimator $\tilde{\theta}^*$ from the true SNR. At first glance, the optimal decision rule as well as the optimal estimator seem equivalent to Bayesian solutions because of the following reason: The term $E^{m_0,m_1}_{\lambda}$ can be interpreted as the posterior probability of $H_1$ given the observations $\{b[1], \ldots, b[m]\}$ and the prior $\lambda(\rho)$ that has been scaled by a cost coefficient. Similarly, $\forall i \geq 0$, the terms $E^{m_0,m_1}_{\mu,i}$ can be interpreted as the conditional moments of the posterior distribution of $\rho$ (or, $\theta$) with prior $\mu$, and the optimal estimator $\tilde{\theta}^*$ as the posterior expected value of $\theta-2$.

However, the proposed scheme is not equivalent to the Bayesian procedure. Note that, $\lambda$ and $\mu$ both behave as
``priors'' for $\rho$, but can be chosen \emph{independently}. For the proposed approach to be Bayesian, one would
require $\lambda = \mu$. In the case of a single constraint under either hypothesis, the corresponding Lagrange
multiplier can always be interpreted as a prior density scaled by a cost coefficient, so that the optimal method
necessarily has a Bayesian equivalent; compare the classic likelihood-ratio test \cite{Lehmann2005}. In our approach,
however, the two Lagrange multipliers $\lambda$ and $\mu$ correspond to two different constraints under the same
distribution, so that there is no equivalence to the Bayesian setup.

Returning to the solution of \eqref{eq:optimal_stopping}, the main advantage of transforming Gaussian sequences into Bernoulli sequences is that the pair $(m_0,m_1)$ becomes a sufficient statistic of $\{b[1], \dots, b[m]\}$. In comparison, directly solving the problem in the SNR-domain requires a more complicated sufficient statistic, which will include observations from $\bm{x}$ and $\tilde{\bm{w}}$ as well as their empirical variances. Given a maximum sample number $\bar{m}$, the matrices $\bm{R}, \bm{G} \in \mathbb{R}^{\bar{m} \times \bar{m}}$ can be calculated via backward recursion with starting point $G_{\bar{m},\bar{m}} = R_{\bar{m},\bar{m}}$. Since the state variables are integers, this recursion is numerically stable for moderately large values of $m_0$ and $m_1$. The final element of the recursion is $R_{0,0}$, which is the cost at the beginning of the test ({\ie}, $m_0 = m_1 = 0$) when the optimal decision and stopping rules, given by \eqref{eq:decision_rule} and \eqref{eq:optimal_stop}, respectively, are used and $\lambda$ and $\mu$ are given, {\ie}, $L(\lambda,\mu) = R_{0,0}(\lambda,\mu)$.
Once we are able to evaluate $L(\lambda,\mu)$, the Lagrange multiplies can be determined by solving \eqref{eq:lagrange_dual} for $\lambda$ and $\mu$. Since by construction $L(\lambda,\mu)$ is jointly concave in $\lambda$ and $\mu$, this is a convex optimization problem that can be solved using standard algorithms, as shown in the next section.

\vspace{-0.1in}
\section{Experimental Results}\label{sec:experiments}\vspace{-0.1in}
In this section, we present results of two numerical experiments. which provide interesting insights into the structure
of the joint detection and estimation problem. The SNR is chosen in the range $[-3\,\text{dB}, 10\,\text{dB}]$ with
grid points at integer values, {\ie}, $\theta_k = 10^{\frac{k-4}{10}}$, $k=1, \dots, 14$. The nominal SNR value
$\theta^*$ is set at $3\,$dB. The target error probabilities are identical for both experiments, with $\alpha =
\beta(\theta) = 0.05$. As a measure for the estimation accuracy, the relative (or normalized) MSE is used and bounded
by a constant. In order to match the problem formulation in \secref{sec:problem_formulation}, the constraint can be
expressed in terms of the absolute MSE and an SNR dependent bound $\gamma(\theta) = c \, \theta^2$ or, in terms of
$\rho$, as $\gamma(\rho) = c \, (1/\rho-2)^2$. For the first experiment, $c = 0.1$, while for the second $c = 0.25$.

For the numerical solution of \eqref{eq:lagrange_dual}, we employed the Subplex algorithm \cite{Rowan1990} as implemented in \cite{Johnson}. It applies the Nelder--Meat simplex algorithm \cite{Nelder1965} in a repeated fashion on suitably chosen low-dimensional subspaces and does not require the calculation of gradients, which is computationally intensive for the recursively-defined cost function \eqref{eq:lagrange_dual}. By limiting the search to subspaces, the algorithm can exploit the sparsity in the optimal Lagrange multipliers. Since there is no formal proof of convergence, the Subplex solution was subsequently verified by evaluating its first-order optimality conditions. The maximum number of Bernoulli samples was set to $\bar{m} = 400$, which proved to be sufficient.

\begin{figure}[htb]
\centering
\includegraphics{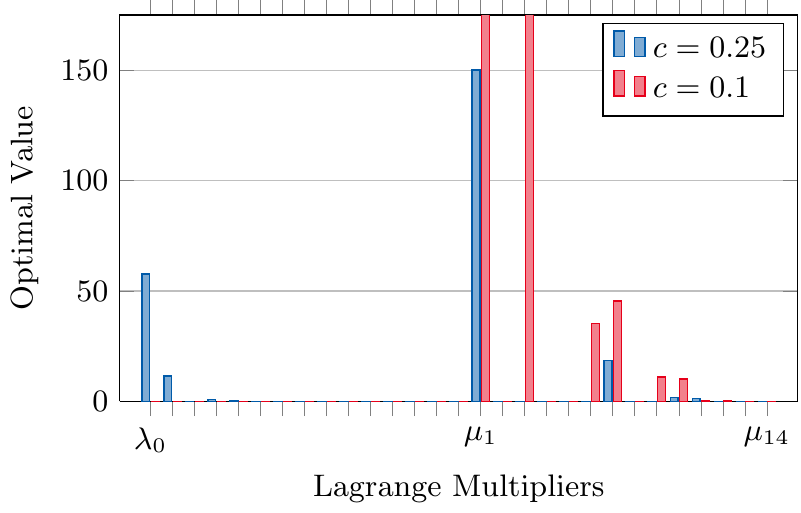}
\caption{Optimal Lagrange multipliers for $\alpha = \beta(\theta) = 0.05$ and $c \in \{0.1, 0.25\}$. The multipliers are plotted in the order $\lambda_0, \dots,\lambda_{14}, \mu_1, \dots,\mu_{14}$. The two clipped values are $\mu_1 = 502.77$ and $\mu_3 = 307.37$.}
\label{fig:lagrange_multipliers}
\end{figure}
\begin{figure}[!t]
  \centering
  \subfloat{\includegraphics{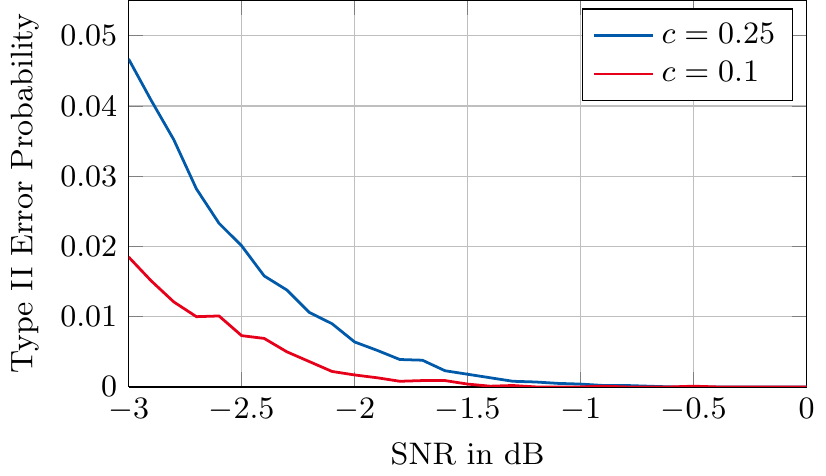}} \\
  \subfloat{\includegraphics{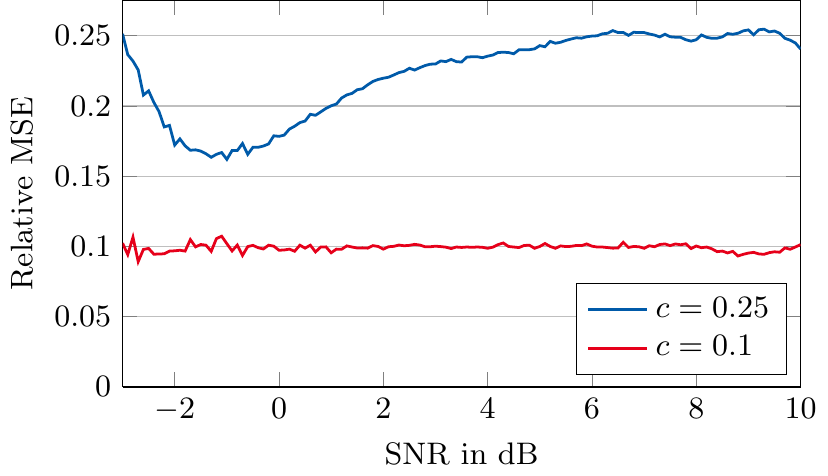}} \\
  \caption{Detection and estimation performance of the proposed algorithm for $\alpha =
           \beta(\theta) = 0.05$ and $c \in \{0.1, 0.25\}$.}
  \label{fig:performance}
\end{figure}
In \figref{fig:lagrange_multipliers}, the optimal Lagrange multipliers are depicted for both experiments. It is
interesting to note that in both cases the solution is sparse, which implies that most of the performance constraints
are inactive. Considering the very coarse SNR grid, this outcome is rather unexpected and suggests that in practice
very few constrains can be sufficient to bound the performance over large SNR intervals. This can also be seen in
\figref{fig:performance}, where the type II error probabilities and the relative MSE are plotted over the range of
SNRs. The results were obtained by averaging over $10^4$ Monte Carlo simulations and the SNR interval was sampled at
intervals of size $0.1\,$dB. Within the numerical accuracy, the performance requirements are met or exceeded for all
SNR values in the feasible interval. Especially the type II error probabilities are well below the required $5\%$ for
all SNR values. For $c = 0.1$, the estimation constraint is so much tighter than the detection constraint that the
latter is virtually deactivated, {\ie}, the corresponding Lagrange multipliers are close to zero (see
\figref{fig:lagrange_multipliers}). The constraints on the relative MSE are stricter so that the bound is reached over
large regions of the SNR interval. Considering that the performance is restricted to integer values of the SNR, it is
remarkable that the requirements are satisfied over the entire interval.

\begin{figure}[h!]
  \centering
  \includegraphics{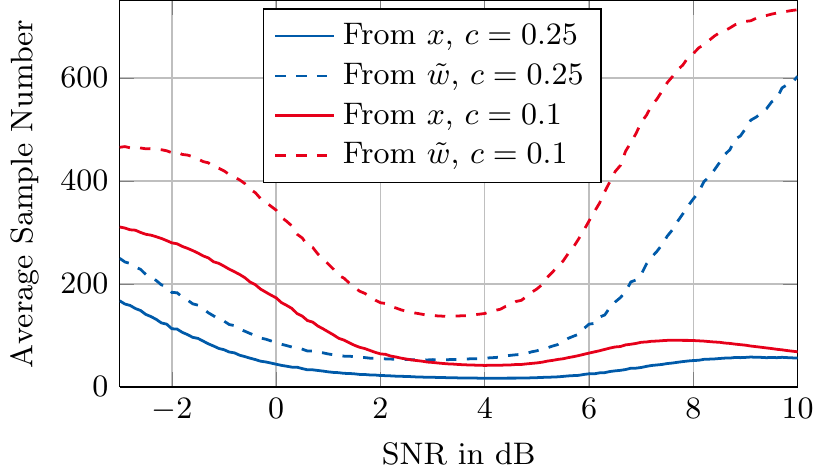}
  \caption{The average number of samples drawn from the observation reference sequences versus SNR.}
  \label{fig:asn}
\end{figure}
The average number of samples drawn from the observation sequence $\bm{x}$ and the reference sequence $\tilde{\bm{w}}$
is shown in \figref{fig:asn}. As expected, ASNs for both cases are high for low SNRs and decrease for higher SNRs. Both
ASNs reach a minimum at around $3\,$dB, which corresponds to the nominal SNR that was targeted in the minimization
procedure. For large SNR values, the number of samples drawn from the reference sequence increases again, while the
number of samples generated by the signal itself stays almost constant. This is a desirable property, considering that
generating training samples is usually easier than taking physical measurements. The modes at low and high SNR values
are due the detection and estimation constraints, respectively.

\vspace{-0.15in}
\section{Concluding remarks}\label{sec:conclusion}\vspace{-0.1in}
We have considered the problem of joint signal detection and SNR estimation for a linear Gaussian model in a sequential
framework. The central idea of our approach is to transform the observed sequences to a sequence of Bernoulli random
variables. This transformation leads to a simpler reformulation of the main optimization problem, which can be
efficiently solved. The expected minimum number of samples required to achieve the desired performance remains almost
constant for increasing values of SNR. We also obtain a sufficient statistic for the test which is very easy to
compute. Experimental results indicate that many constraints on the optimization setup are inactive, which renders the
problem easily solvable. These results indicate the feasibility of the proposed method to practical applications.
Understanding the implications of non-stationary noise processes on the performance of the proposed approach especially
in the high-SNR regime is one of the main avenues for future research.

\bibliographystyle{IEEEbib}
\bibliography{icassp2017}

\end{document}